\documentclass[conference]{IEEEtran}
\IEEEoverridecommandlockouts

\usepackage{amsmath,amssymb,amsfonts}
\usepackage{cite}
\usepackage{graphicx}
\usepackage{textcomp}
\usepackage{xcolor}
\usepackage{listings}
\usepackage{subfig}
\usepackage{pdfpages}
\usepackage{hyperref}
\usepackage{tcolorbox}
\graphicspath{ {./images/} }
\definecolor{yamlGreen}{RGB}{63,127,63}
\definecolor{yamlBlue}{RGB}{0,0,255}
\definecolor{yamlPurple}{RGB}{127,0,127}
\definecolor{yamlGray}{RGB}{100,100,100}

\lstdefinelanguage{yaml}{
  keywords={true,false,null,y,n,yes,no,on,off},
  keywordstyle=\color{yamlBlue}\bfseries,
  basicstyle=\ttfamily\small,
  sensitive=false,
  comment=[l]{\#},
  commentstyle=\color{yamlGreen}\itshape,
  stringstyle=\color{yamlPurple},
  morestring=[b]',
  morestring=[b]",
  moredelim=[l][\color{yamlGray}]:-,
  moredelim=[l][\color{yamlGray}]:,
  moredelim=*[l][\color{yamlGray}]{-},
  moredelim=**[is][\color{yamlGray}]{:}{\ },
  classoffset=0,
  otherkeywords={>, -, \{, \}, \[, \], :},
  morekeywords={[2]0,1,2,3,4,5,6,7,8,9},
  keywordstyle={[2]\color{yamlBlue}},
  literate=
    {>}{{\textcolor{yamlGray}{>}}}1
    {-}{{\textcolor{yamlGray}{-}}}1
    {:}{{\textcolor{yamlGray}{:}}}1,
}

\def\BibTeX{{\rm B\kern-.05em{\sc i\kern-.025em b}\kern-.08em
    T\kern-.1667em\lower.7ex\hbox{E}\kern-.125emX}}
\begin{document}

\title{LLMSecConfig: An LLM-Based Approach for Fixing Software Container Misconfigurations}

\author{
\IEEEauthorblockN{Ziyang Ye\IEEEauthorrefmark{1}\IEEEauthorrefmark{2}, 
Triet Huynh Minh Le\IEEEauthorrefmark{1}\IEEEauthorrefmark{2} 
and M. Ali Babar\IEEEauthorrefmark{1}\IEEEauthorrefmark{2}} 
\IEEEauthorblockA{\IEEEauthorrefmark{1}\textit{CREST - The Centre for Research on Engineering Software Technologies}, Adelaide, Australia\\ 
\IEEEauthorblockA{\IEEEauthorrefmark{2}\textit{School of Computer and Mathematical Sciences}, \textit{The University of Adelaide}, Adelaide, Australia\\ 
ziyang.ye@student.adelaide.edu.au,
triet.h.le@adelaide.edu.au, ali.babar@adelaide.edu.au}}
}

\maketitle

\begin{abstract}
Security misconfigurations in Container Orchestrators (COs) can pose serious threats to software systems. While Static Analysis Tools (SATs) can effectively detect these security vulnerabilities, the industry currently lacks automated solutions capable of fixing these misconfigurations. The emergence of Large Language Models (LLMs), with their proven capabilities in code understanding and generation, presents an opportunity to address this limitation. This study introduces LLMSecConfig, an innovative framework that bridges this gap by combining SATs with LLMs. Our approach leverages advanced prompting techniques and Retrieval-Augmented Generation (RAG) to automatically repair security misconfigurations while preserving operational functionality. Evaluation of 1,000 real-world Kubernetes configurations achieved a 94\% success rate while maintaining a low rate of introducing new misconfigurations.

Our work makes a promising step towards automated container security management, reducing the manual effort required for configuration maintenance.
\end{abstract}


\begin{figure*}[!htb]
\centering
\begin{minipage}[t]{0.45\linewidth}
\begin{lstlisting}[language=yaml,
    numbers=none,
    escapeinside={(*}{*)},
    basicstyle=\ttfamily\scriptsize]
apiVersion: v1
kind: Pod
metadata:
  name: <Pod name>
spec:
  containers:
  - name: <container name>
    image: <image>
    securityContext:
      allowPrivilegeEscalation: (*\textcolor{red}{true}*)
\end{lstlisting}
\end{minipage}%
\hspace{0.05\linewidth}%
\begin{minipage}[t]{0.45\linewidth}
\begin{lstlisting}[language=yaml,
    numbers=none,
    escapeinside={(*}{*)},
    basicstyle=\ttfamily\scriptsize]
apiVersion: v1
kind: Pod
metadata:
  name: <Pod name>
spec:
  containers:
  - name: <container name>
    image: <image>
    securityContext:
      allowPrivilegeEscalation: (*\textcolor{green}{false}*)
\end{lstlisting}
\end{minipage}
\caption{Comparison of misconfigured (left) versus properly configured (right) Kubernetes security contexts. The misconfigured configuration allows privilege escalation, enabling container processes to gain elevated privileges beyond their parent processes. In contrast, the secure configuration explicitly disables this capability.}
\label{fig:k8s-security-comparison}
\end{figure*}

\section{Introduction}
Container Orchestrators (COs) have become integral to modern cloud infrastructure, enabling scalable and efficient deployment of applications. However, their widespread adoption has also highlighted a critical security challenge: the prevalence of security misconfigurations. These misconfigurations are the leading cause of vulnerabilities in containerised environments, posing significant risks to system integrity and data confidentiality \cite{casalicchio2019container, shamim2020xi}. While Static Analysis Tools (SATs) are proficient at detecting these vulnerabilities through pattern matching and rule-based analysis \cite{pistoia2007survey}, there remains a substantial gap in automated remediation solutions. This gap necessitates manual intervention by security administrators, a process that is not only time-consuming and labour-intensive but also prone to human error, especially as the scale and complexity of deployments grow \cite{10173942, 9476056,russell2024centralizeddefenseloggingmitigation}.

The manual process of fixing misconfigurations involves meticulous examination and correction of numerous configuration files, often requiring deep domain expertise and an in-depth understanding of interdependent settings. This complexity is exacerbated in dynamic environments where configurations frequently change due to scaling operations and Continuous Integration and Continuous Deployment (CI/CD) practices~\cite{arani2024systematic}. There are numerous changes in adding, removing, and modifying different services, along with a large number of corresponding configuration files in CI/CD pipelines. Consequently, organizations usually face significant operational bottlenecks and increased vulnerability exposure due to the lag between the detection and remediation of security issues (misconfigurations) in COs.

Advancements in Deep Learning and, recently, Large Language Models (LLMs) in Software Engineering offer a promising avenue to bridge this gap~\cite{le2020deep,hou2024large}. LLMs have demonstrated remarkable capabilities in understanding and generating complex technical content \cite{bhatt2023purplellamacybersecevalsecure, bhatt2024cyberseceval2widerangingcybersecurity, wan2024cyberseceval3advancingevaluation}, making them suitable candidates for automating the remediation of security misconfigurations. By leveraging LLMs in conjunction with Retrieval-Augmented Generation (RAG) techniques, it is possible to not only identify but also generate accurate and context-aware fixes for misconfigurations, thereby significantly reducing the manual effort and minimising the risk of introducing new vulnerabilities.

This study introduces LLMSecConfig, a novel framework designed to automate the repair of security misconfigurations in COs, ensuring that configurations adhere to security best practices while maintaining operational functionality.

This paper makes the following key contributions:
\begin{enumerate}
    \item \textbf{Novel Framework for Automated Security Configuration Repair}: An innovative framework that integrates SATs with LLMs for automated security configuration repair in COs. Our approach uniquely combines LLMs with RAG techniques to generate precise fixes while maintaining operational functionality.
    
    \item \textbf{Demonstrated Effectiveness with Real-world Evaluation}: Comprehensive evaluation using 1,000 real-world Kubernetes configurations, demonstrating superior performance with a 94.3\% repair success rate and minimal error introduction (0.024). Our approach significantly outperforms baseline models, including GPT-4o-mini, which achieved only a 40.2\% success rate.
    
    \item \textbf{Open-source Dataset and Implementation}: To facilitate reproducibility and future research in container security, we provide our complete implementation, including the collected dataset of Kubernetes misconfiguration, source code of the framework, and evaluation scripts at~\url{https://figshare.com/s/2a9be8ccfbec9d8ba199}. 
\end{enumerate}

\section{Background and Motivation}
\subsection{Container Security Landscape}
The rise of Infrastructure as Code (IaC) has transformed IT infrastructure management, enabling automated and scalable deployment of applications. Container technologies like Kubernetes offer significant benefits in terms of portability and resource efficiency, but they also present unique security challenges that span multiple layers of the system. These challenges range from kernel vulnerabilities to intricate orchestration misconfigurations \cite{Casalicchio2020TheSI, yang2021security, sultan2019container}. 

Fig. \ref{fig:k8s-security-comparison} illustrates a common security misconfiguration in container orchestration, where incorrectly enabled privilege escalation could compromise system security. However, security fixes in container orchestrators are often complex and go beyond simple value correction. For instance, setting a Pod's \textit{allowPrivilegeEscalation} to \textit{false} becomes challenging when multiple security configurations interact across namespaces.

Users must analyse configuration dependencies and prevent privilege overrides by considering three main factors. Primarily, \textit{CAP\_SYS\_ADMIN} will override \textit{allowPrivilegeEscalation} to \textit{true}, invalidating the previous setting. Consequently, any predefined \textit{CAP\_SYS\_ADMIN} must be removed to ensure that \textit{allowPrivilegeEscalation} remains \textit{false}. This behaviour is notably different from the scenario shown in Fig. \ref{fig:k8s-security-comparison}.

Additionally, security contexts in Kubernetes operate at different levels. \textit{PodSecurityContext} manages pod-level security and shared container settings, while container-level \textit{SecurityContext} controls individual container security and takes precedence when both specify the same field. This complexity is further compounded by varying behaviours between Linux and Windows containers.

Moreover, security configuration complexity scales significantly with the architectural hierarchy of Kubernetes clusters. At the highest level, a cluster encompasses multiple nodes. Each node, in turn, hosts numerous Pods, while individual Pods can run multiple applications. Throughout this hierarchy, components maintain their own security settings and network policies, creating an intricate web of security configurations.

This hierarchical structure makes manually defining and fixing rules extremely challenging. More importantly, incomplete fixes can introduce new issues, such as incompatible values between interdependent configurations. Therefore, addressing these intricate security challenges requires automated and systematic approaches rather than manual intervention.

Modern COs provide an extensive array of configuration options to support diverse deployment scenarios and security requirements. While this flexibility is beneficial, it significantly increases the complexity of maintaining secure settings across all components \cite{truyen2019comprehensive}. The dynamic nature of container environments, characterised by frequent updates and scaling operations, further complicates the security landscape by potentially introducing vulnerabilities through configuration changes \cite{casalicchio2019container}.

Multi-tenancy environments present additional security challenges, where shared resources and namespaces require careful management to prevent unauthorised access and resource contention \cite{sultan2019container}. Supply chain security is equally critical, as container images and their dependencies can introduce vulnerabilities that demand continuous monitoring \cite{yang2021security}. The rapid evolution of security threats, including side-channel attacks and container-specific exploits, necessitates robust and adaptable security measures \cite{sultan2019container, bose2021under}.

\subsection{The Need for Automated Fixing of Software Container Misconfigurations}
The complexity of container security configurations, combined with the rapid pace of modern development practices, creates a compelling need for automated remediation solutions. In large-scale deployments, the number of configuration files can reach into thousands, making manual corrections time-consuming and prone to errors \cite{10173942, 9476056, russell2024centralizeddefenseloggingmitigation}. Security administrators must address each misconfiguration meticulously to ensure consistency and compliance across the entire infrastructure. As deployments grow, the operational burden and response times increase, rendering manual approaches impractical.

Currently, it is challenging to address misconfigurations in COs like Kubernetes.
Firstly, understanding the intricate relationships between different settings and ensuring that fixes do not conflict with existing configurations requires significant expertise and domain knowledge \cite{haque2022kgsecconfig}. This complexity makes effective manual remediation challenging, especially in environments lacking specialised security personnel.

The rapid pace of modern development practices, such as CI/CD, also demands swift and reliable configuration updates to keep up with agile development cycles and frequent deployments. Manual remediation cannot meet the speed and efficiency required by these practices, leading to security risks and deployment delays \cite{10173942, 9476056, russell2024centralizeddefenseloggingmitigation}.
Furthermore, regulatory standards often mandate stringent security configurations and require detailed audit trails of configuration changes. Automated fixing can ensure consistent adherence to compliance requirements and provide comprehensive logs for auditing purposes \cite{shevchuk2023softwarefi}.

Despite the clear need for automated remediation solutions, current research and industry practices have not adequately addressed the automation of security misconfiguration fixes in container environments. While significant progress has been made in automated detection through SATs and validation frameworks, the critical step of actually implementing corrections remains largely manual\cite{pistoia2007survey}. Existing approaches primarily focus on providing guidance and recommendations, leaving the complex task of modification and validation to human operators\cite{4534479, 6987569, 10605161, al2024containers}. This limitation is further exacerbated by the heterogeneous nature of COs and SATs; their diverse design philosophies, programming languages, and technical implementations make it challenging to develop a universal framework for automated detection and repair. Consequently, this gap between detection and remediation represents a significant bottleneck in container security management, particularly as organizations scale their deployments. Our work addresses this fundamental limitation by proposing one of the first automated solutions for fixing container security misconfigurations.

\section{LLMSecConfig: An Automated Framework for Fixing Software Container Misconfigurations}

\subsection{Framework Overview}
\begin{figure*}[!htb]
\centering
\includegraphics[width=1.0\textwidth]{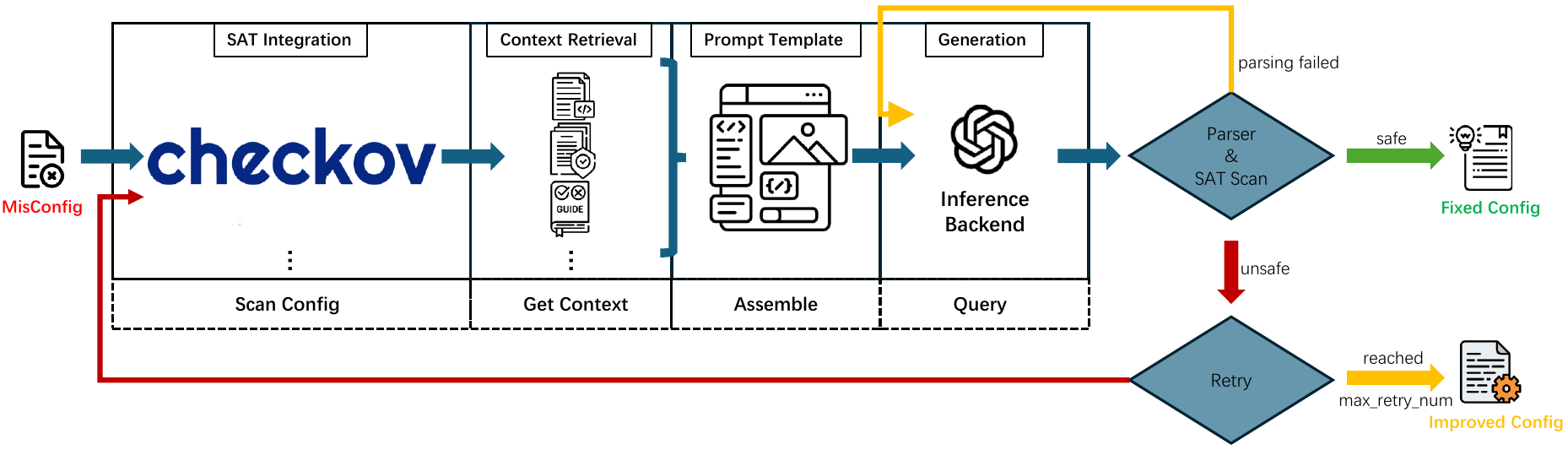}
\caption{Architecture overview of the LLMSecConfig framework for automated Kubernetes security configuration repair. The pipeline comprises three main stages: (1) configuration analysis using SATs, (2) context retrieval through RAG, and (3 \& 4) repair generation using Large Language Models (LLMs). The system iteratively processes configurations until all security issues are resolved or the maximum number of attempts is reached.}
\label{fig:framework}
\end{figure*}

LLMSecConfig presents an end-to-end pipeline designed to detect and rectify misconfigurations in COs by synergizing SATs with LLMs. The framework operates through three interconnected phases: (\textit{i}) SAT Integration and (\textit{ii}) Context Retrieval for detecting and understanding misconfigurations, and then (\textit{iii}) Repair Generation and Validation, as illustrated in Fig. \ref{fig:framework}. This modular architecture ensures flexibility and scalability, allowing each component to function independently while contributing to the overall system's effectiveness.

\subsection{Static Analysis Tools Integration for Misconfiguration Detection}
SATs are instrumental in the identification of security vulnerabilities and misconfigurations within software systems \cite{6987569,le2021large}. Among the various SATs available for COs, such as Checkov, Trivy, Terrascan, and KubeSec \cite{checkov2024, trivy2024, terrascan2024, kubesec2024}, our framework selects Checkov as the primary tool due to its strategic advantages. Checkov offers a comprehensive collection of over 1,000 built-in security policies, ensuring extensive coverage of potential vulnerabilities\cite{checkov2024}. Its implementation in pure Python facilitates seamless integration with our framework, allowing direct access to policy definitions and enhancing flexibility. Additionally, Checkov provides rich contextual information through integration with Prisma Cloud documentation and receives regular updates that incorporate the latest security best practices and emerging vulnerability patterns\cite{prisma_cloud}.

\subsection{Context Retrieval for Misconfiguration Understanding}
The context retrieval module is a critical component of LLMSecConfig, where RAG, with its efficient retrieval capability, has the characteristics of knowledge augmentation and the performance of context-aware generation \cite{gao2023retrieval}, which enables us to enhance the repair generation process. This module integrates corrective feedback mechanisms to elevate the quality and reliability of the fixes produced by the LLMs \cite{yan2024corrective}. For every detected misconfiguration, the system aggregates three essential types of information to provide a comprehensive context for repair.

Firstly, the SAT output serves as the foundational context, offering detailed error messages from Checkov scans, specific policy violation identifiers (e.g., CKV\_K8S\_20), resource type and location information, links to relevant documentation, and severity levels within the framework context, as illustrated in Fig. \ref{lst:checkov-output}. This information provides a precise understanding of the nature and location of each misconfiguration.

\begin{figure}[t]
    \begin{minipage}{1.0\columnwidth}
        \begin{lstlisting}[
            basicstyle=\ttfamily\scriptsize,
            breaklines=true,
            breakatwhitespace=false,
            linewidth=\linewidth,
            numbers=none,
            escapechar=|
        ]
$ checkov -f your_input.yaml --compact --framework kubernetes --quiet -c CKV_K8S_20
|\textcolor{blue}{kubernetes scan results:}|
|\textcolor{cyan}{Passed checks: 0, Failed checks: 1, Skipped checks: 0}|
Check: CKV_K8S_20: "Containers should not run with allowPrivilegeEscalation"
|\textcolor{red}{FAILED for resource: Pod.default.\{pod-name\}}|
|\textcolor{magenta}{File: /your\_input.yaml:1-10}|
Guide: https://docs.prismacloud.io/en/enterprise-edition/policy-reference/kubernetes-policies/kubernetes-policy-index/bc-k8s-19
        \end{lstlisting}
    \end{minipage}
    \caption{Example output from the Checkov security scanner, indicating a detected privilege escalation vulnerability in a Kubernetes configuration.}
    \label{lst:checkov-output}
\end{figure}

Secondly, the policy implementation source code, exemplified in Fig. \ref{lst:checkov-sourcecode}, offers in-depth insights into the validation logic and conditions underpinning each security policy, including the complete Python implementation of security checks, key configuration parameters being evaluated, and expected values. Access to the source code allows the framework to comprehend the specific criteria used to determine compliance, enabling more accurate and context-aware repairs. For each ready-to-fix configuration, we first scan it with the Checkov tool to identify security issues and obtain detailed information. Each security issue has a unique ID in Checkov, and by examining its GitHub repository\cite{checkov2024}, we can locate the corresponding Python file for that ID, which includes the implementation of the security-related checking policy. To facilitate fast indexing, we created a CSV file to map each ID to its source code file.

\begin{figure}[!htb]
    \begin{minipage}{1.0\columnwidth}
        \begin{lstlisting}[
            language=Python,
            basicstyle=\ttfamily\scriptsize,
            breaklines=true,
            breakatwhitespace=false,
            numbers=none,
            linewidth=\linewidth
        ]
from typing import Dict, Any
from checkov.common.models.enums import CheckResult
from checkov.kubernetes.checks.resource.base_container_check \
    import BaseK8sContainerCheck
class AllowPrivilegeEscalation(BaseK8sContainerCheck):
    def init(self) -> None:
        # CIS-1.3 1.7.5
        # CIS-1.5 5.2.5
        name = "Containers should not run with allowPrivilegeEscalation"
        id = "CKV_K8S_20"
        super().init(name=name, id=id)
    def scan_container_conf(self, metadata: Dict[str, Any], conf: Dict[str, Any]) -> CheckResult:
        self.evaluated_container_keys = [
            "securityContext/allowPrivilegeEscalation"
        ]
        if isinstance(conf, dict) and conf.get("securityContext"):
            if conf["securityContext"].get(
                "allowPrivilegeEscalation"
            ) is False:
                return CheckResult.PASSED
        return CheckResult.FAILED
check = AllowPrivilegeEscalation()
        \end{lstlisting}
    \end{minipage}
    \caption{Python implementation of the Checkov security check for container privilege escalation.}
    \label{lst:checkov-sourcecode}
\end{figure}

\begin{figure}[!htb]
    \begin{minipage}[b]{1.0\columnwidth}
        \fbox{\includegraphics[width=\columnwidth]{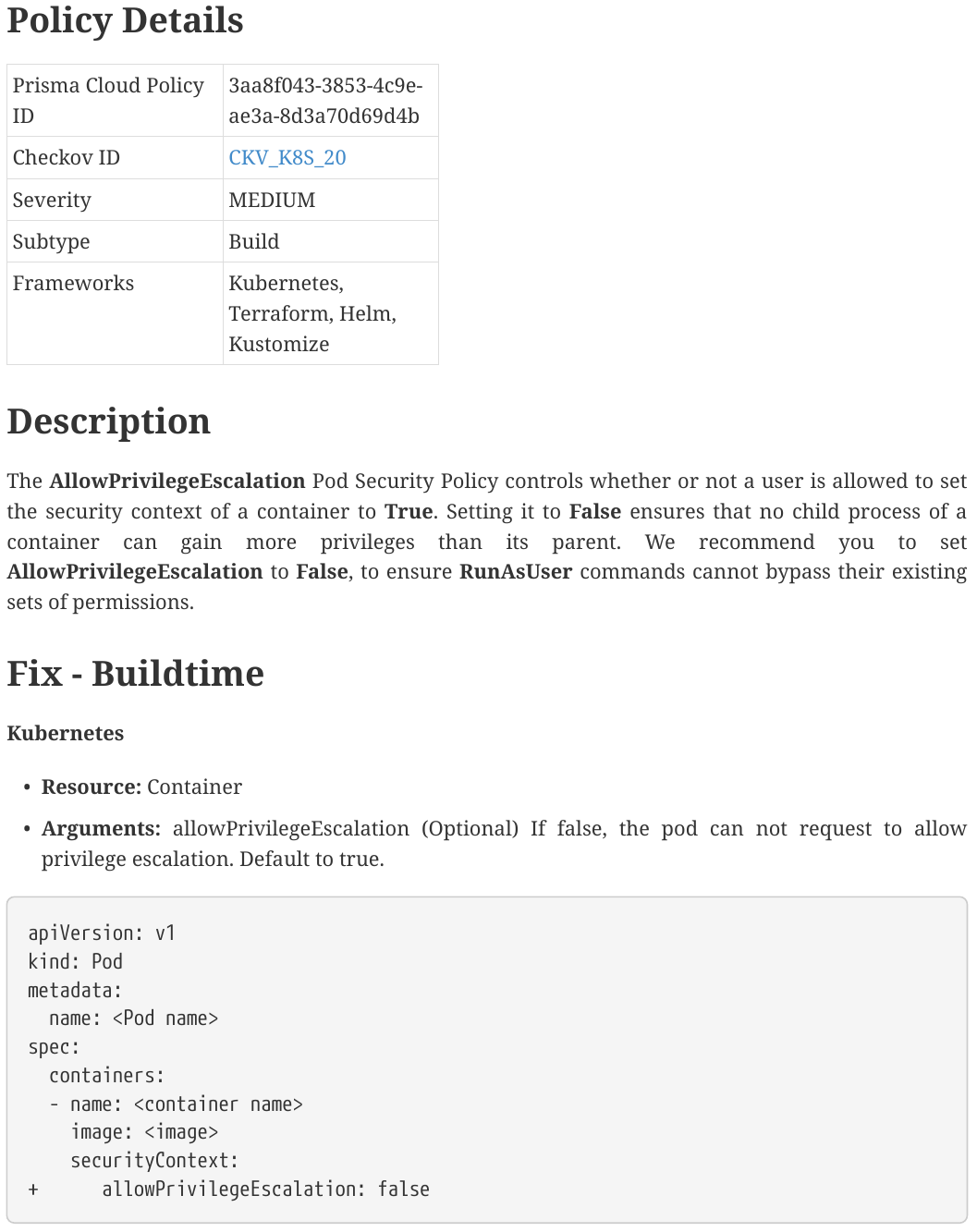}}
    \end{minipage}
    \caption{Prisma Cloud security policy documentation for CKV\_K8S\_20: \textbf{Containers Run with AllowPrivilegeEscalation}.}
    \label{fig:prisma}
\end{figure}

Thirdly, comprehensive security documentation from Prisma Cloud, depicted in Fig. \ref{fig:prisma}, supplements the context with detailed analysis of security implications, specific fix recommendations, example configurations demonstrating correct implementations, and framework-specific guidance and best practices. This documentation provides additional layers of information that guide the LLMs in generating precise and effective fixes.

The context module efficiently retrieves and maintains the relationships between these information sources by extracting policy identifiers and error details from Checkov's output, locating the corresponding Python implementations, and fetching the associated Prisma Cloud documentation. The merge of these elements ensures that the LLMs can get all the necessary information to generate accurate and contextually appropriate repairs. More details about the prompt templates can be found in the Appendix at \url{https://figshare.com/s/2a9be8ccfbec9d8ba199}.

\subsection{Misconfiguration Repair Generation and Validation}

The repair generation component leverages LLMs to produce context-aware fixes for the identified misconfigurations. This process involves carefully designed prompt templates that incorporate error information, policy implementation details, and security documentation. By providing the LLM with this rich context, we enable it to generate precise and effective repairs. The system supports multiple LLM backends, offering flexibility in model selection based on specific requirements.

Upon receiving the context, the LLM generates proposed configuration changes to resolve the identified security issues. To ensure that these generated fixes are valid and effective, LLMSecConfig incorporates a validation process. This process begins with using SAT to verify valid syntax, ensuring that the generated output maintains a valid YAML structure. If the syntax is valid, the system proceeds to security validation, confirming that the repairs address the identified vulnerabilities without introducing new issues.

To enhance system robustness, we implement two stages of retry mechanisms. An outer loop manages the overall repair attempts, limiting the total number of fixes attempted for each configuration to prevent endless loops. An inner loop handles potential parsing failures of LLM outputs, ensuring that only valid YAML configurations proceed to the security validation stage.

During each repair iteration, the system performs SAT analysis to identify any remaining security issues. If issues persist, the context is updated, and the LLM generates a new fix. To conform to specific SAT's security best practices, this iterative process continues until all issues are resolved, the maximum number of repair attempts is reached, or an unrecoverable error is encountered. 

Additionally, LLMSecConfig maintains comprehensive control mechanisms and detailed logging. Each repair attempt is logged with information about the original configuration state, identified security issues, generated fixes, and validation results.

By combining automated repair generation with stringent validation protocols, LLMSecConfig provides a reliable approach to addressing security misconfigurations in COs. The integration of SATs and LLMs, coupled with effective context management and validation, ensures that the framework can effectively repair misconfigurations while preserving the operational functionality of the containerised applications.

\section{Experimental Setup and Design}
\subsection{Kubernetes as a Case Study Container Orchestrator}
Kubernetes was selected as our experimental platform due to its position as the de facto standard for container orchestration, with over 92\% of organizations either using or evaluating it according to the CNCF survey \cite{cncf_survey_report_2020, edgedelta_kubernetes_adoption_statistics}. Its extensive security-related settings and well-documented configuration patterns provide an ideal testing ground for our automated repair approach, making our research findings immediately applicable to a large user base \cite{k8s2024}.

\subsection{Dataset}
\label{subsec:dataset}
The need for a custom dataset stems from several key factors. First, there is an absence of public datasets specifically designed for container misconfiguration remediation\cite{Khraisat2021ACR, Mahajan2022DetectionAA}. Moreover, even if such datasets existed, they would require constant updates due to the evolving nature of COs and their security configurations~\cite{le2019automated}. As API specifications change, security configuration requirements must adapt accordingly.

We selected ArtifactHub as our primary data source due to its repository of high-quality, vetted Kubernetes configurations that accurately represent real-world usage patterns. As the official CNCF artifact hub \cite{artifacthub, cncf_artifact_hub}, ArtifactHub provides a standardised and open format, facilitating straightforward data collection and validation processes. ArtifactHub ensures that the configurations are well-maintained and regularly updated, capturing current configuration trends and security challenges.

The data collection process was executed through a systematic multi-step approach. Initially, we accessed ArtifactHub's API to identify the top 1,000 most popular projects based on their internal ranking system, ensuring that our dataset encompassed configurations widely used and representative of common industry practices. Each selected project, managed using Helm, required conversion of Helm charts into raw Kubernetes configuration files. This step was crucial as Helm charts typically generate multiple YAML files, each representing distinct Kubernetes resources or components.

\begin{figure*}[!htb]
    \centering
    \includegraphics[width=1.0\textwidth]{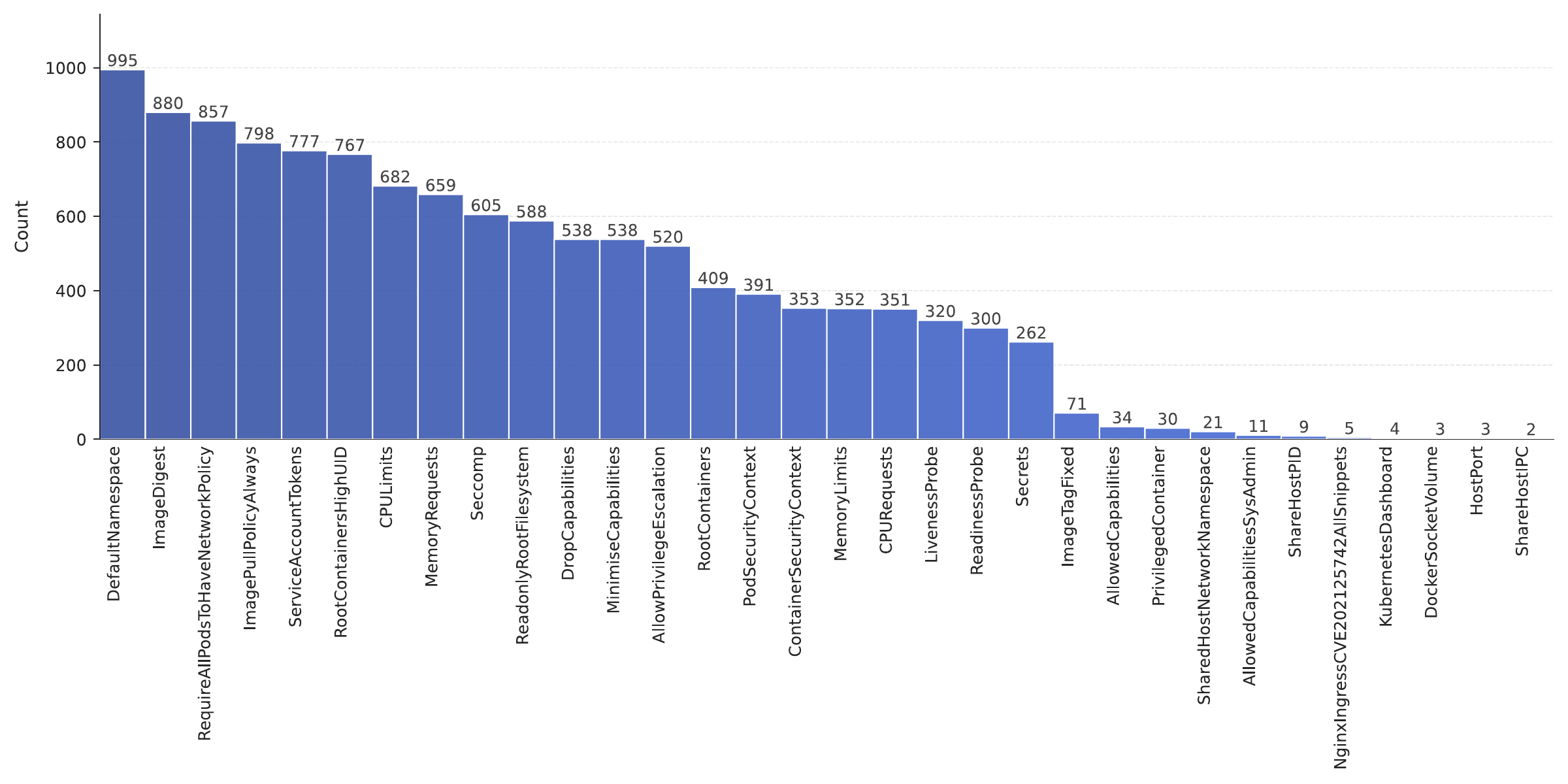}
    \caption{Distribution of security misconfiguration types in our dataset collected from ArtifactHub. The chart illustrates the relative frequency of different security issues detected by SATs across 1,000 configuration files, highlighting the prevalence of specific vulnerability patterns in real-world container configurations.}
    \label{fig:data_distro}
\end{figure*}

Subsequently, the exported YAML files were parsed to isolate individual sub-configurations. Each sub-configuration was treated as a separate entity corresponding to a specific Kubernetes resource, such as Pods, Services, or Deployments. These isolated configurations were then subjected to analysis using SATs to identify security misconfigurations. Only those files that triggered security issues were retained to increase the vulnerability label quality~\cite{croft2023data,le2024automatic}, culminating in a robust test dataset of 1,000 configuration files with detected misconfigurations. The distribution of different types of security misconfigurations within this dataset is illustrated in Fig.~\ref{fig:data_distro}, highlighting the prevalence of various vulnerability patterns in real-world container configurations.

\subsection{LLMs Selection for Repair Generation}
Our evaluation utilised two distinct LLMs: Mistral Large 2 and GPT-4o-mini\cite{mistral_large_enough, openai_gpt_4o_mini}, strategically selected to compare both open-source and closed-source approaches while considering practical deployment factors.

Mistral Large 2 was chosen as our primary open-source model not only for its strong performance characteristics but also for its exceptional cost-effectiveness. Its open-source nature enables complete transparency in deployment and modification, allowing organizations to implement our framework without dependency on external API services. The model's efficient architecture makes it particularly suitable for production deployment, offering high performance at significantly lower computational costs compared to other open-source alternatives\cite{mistral_large_enough}.

GPT-4o-mini was selected as our closed-source baseline for several practical reasons. As OpenAI's most cost-effective model\cite{openai_gpt_4o_mini}, it offers a good balance between performance and operational costs. Its widespread adoption and extensive usage history in the developer community provide valuable benchmarking opportunities. The model's optimisation for speed makes it particularly suitable for rapid configuration analysis and repair tasks, while still maintaining competitive accuracy levels. Although newer models are available, GPT-4o-mini's established track record and substantial community feedback make it reliable for evaluating our framework's performance with commercial solutions.

\subsection{Evaluation Metrics}

We adopted two primary evaluation metrics, i.e., the \textbf{Parse Success Rate (PSR)} and the \textbf{Pass Rate (PR)}. The PSR measures the ability of LLMs to generate syntactically valid YAML configurations, while the PR indicates the proportion of repaired YAML configurations that successfully pass all security checks without errors.

Let $\mathcal{D} = \{d_1, \ldots, d_N\}$ denotes our test dataset, consisting of $N$ configurations.

\begin{equation}
    \mathrm{Parse\ Success\ Rate} = \frac{1}{N} \sum_{i=1}^N \mathbb{I}_{\text{parse}}(d_i) \times 100\%
\end{equation}

\begin{equation}
    \mathrm{Pass\ Rate} = \frac{1}{N} \sum_{i=1}^N \mathbb{I}_{\text{pass}}(d_i) \times 100\%
\end{equation}

\noindent where $\mathbb{I}_{\text{parse}}(d_i)$ and $\mathbb{I}_{\text{pass}}(d_i)$ are indicator functions that equal 1 if the YAML parsing of configuration $d_i$ succeeds and if configuration $d_i$ passes all security checks, respectively, and 0 otherwise. Higher values are preferred for both metrics.

To evaluate repair efficiency, we measured the \textbf{Average Pass Steps (APS)}:

\begin{equation}
    \mathrm{Average\ Pass\ Steps} = \frac{\sum_{i=1}^N s(d_i) \cdot \mathbb{I}_{\text{pass}}(d_i)}{\sum_{i=1}^N \mathbb{I}_{\text{pass}}(d_i)}
\end{equation}

\noindent where $s(d_i)$ is the number of steps required for configuration $d_i$ to pass all security checks. Lower values are preferred.

For step-wise analysis, we computed the \textbf{Area Under Curve (AUC)} metrics for both \textbf{Pass Rate over Steps ($\mathrm{AUC_{PRS}}$)} and \textbf{Average Pass Steps over Steps ($\mathrm{AUC_{APSS}}$)}:

\begin{equation}
    \mathrm{AUC_{PRS}} = \frac{1}{T} \sum_{t=1}^T \text{PR}_t
\end{equation}

\begin{equation}
    \mathrm{AUC_{APSS}} = \frac{1}{T} \sum_{t=1}^T \text{APS}_t
\end{equation}

\noindent where $T$ is the maximum number of allowed repair steps, $\text{PR}_t$ is the pass rate achieved within $t$ steps, and $\text{APS}_t$ is the average number of steps needed for passing configurations when limited to $t$ steps. Higher values are preferred for $\mathrm{AUC_{PRS}}$, while lower values are preferred for $\mathrm{AUC_{APSS}}$.

Finally, we assessed repair quality through the \textbf{Security Improvement} and \textbf{Average Introduced Errors} metrics:

\begin{equation}
    \mathrm{Security\ Improvement} = \frac{\sum_{i=1}^N \left( e_{\text{init}}(d_i) - e_{\text{final}}(d_i) \right)}{\sum_{i=1}^N e_{\text{init}}(d_i)}
\end{equation}

\noindent where $e_{\text{init}}(d_i)$ represents the number of security issues in configuration $d_i$ before repair, and $e_{\text{final}}(d_i)$ represents the number of security issues in configuration $d_i$ after repair. Higher values are preferred.

\begin{equation}
    \mathrm{Average\ Introduced\ Errors} = \frac{1}{N} \sum_{i=1}^N e_{\text{new}}(d_i)
\end{equation}

\noindent where $e_{\text{new}}(d_i)$ represents the number of new security issues introduced during the repair of configuration $d_i$. Lower values are preferred.

\section{Research Questions and Experimental Results}
\subsection{\textbf{RQ1: To what extent are LLMs effective to correct misconfigurations?}}
\textbf{Motivation:} Our primary research question aims to evaluate the fundamental capability of LLMs in automating configuration error correction. While SATs excel at detecting misconfigurations, their inability to automatically correct them creates a significant operational gap. This question quantitatively assesses whether LLMs can effectively bridge this gap by examining several key aspects. We aim to evaluate correction accuracy rates across different LLMs and analyse the syntax validity of the generated corrections.

\textbf{Method:} We conducted the experiments using the GPT-4o-mini~\cite{openai_gpt_4o_mini} and mistral-large-2407~\cite{mistral_large_enough} models on a dataset of 1,000 Kubernetes configurations as described in Section~\ref{subsec:dataset}. For both models, we set the temperature to 0.5 and used the default values for all other hyperparameters. Additionally, we configured the maximum parser retry time and the maximum retry value to 10 and 5, respectively. The key parameters in our framework, namely \textit{temperature} and \textit{max\_retry}, were selected as follows:

\begin{itemize}
    \item \textbf{\textit{temperature}}: This parameter controls the randomness of the model's output. A value of 0 results in deterministic outputs, while higher values increase variability. We set the temperature to 0.5, as it provides a balance between diversity and stability, based on findings from prior research \cite{10603677}.

    \item \textbf{\textit{max\_retry}}: To account for the non-deterministic nature of LLMs, we introduced a retry mechanism. A non-zero \textit{max\_retry} value increases the likelihood of generating correct outputs, but excessive values can lead to higher computational costs. Through experimentation, we determined that a maximum parser retry time of 10 and a maximum retry value of 5 offer a practical balance between performance and efficiency.
\end{itemize}

For each sample, we logged the input file's Checkov output and the output file's scanning result, so we could obtain all the metrics we designed and the stopping step.

\begin{figure}[t]
    \centering
    \subfloat[Pass Rate over Steps]{
    \includegraphics[width=0.5\columnwidth]{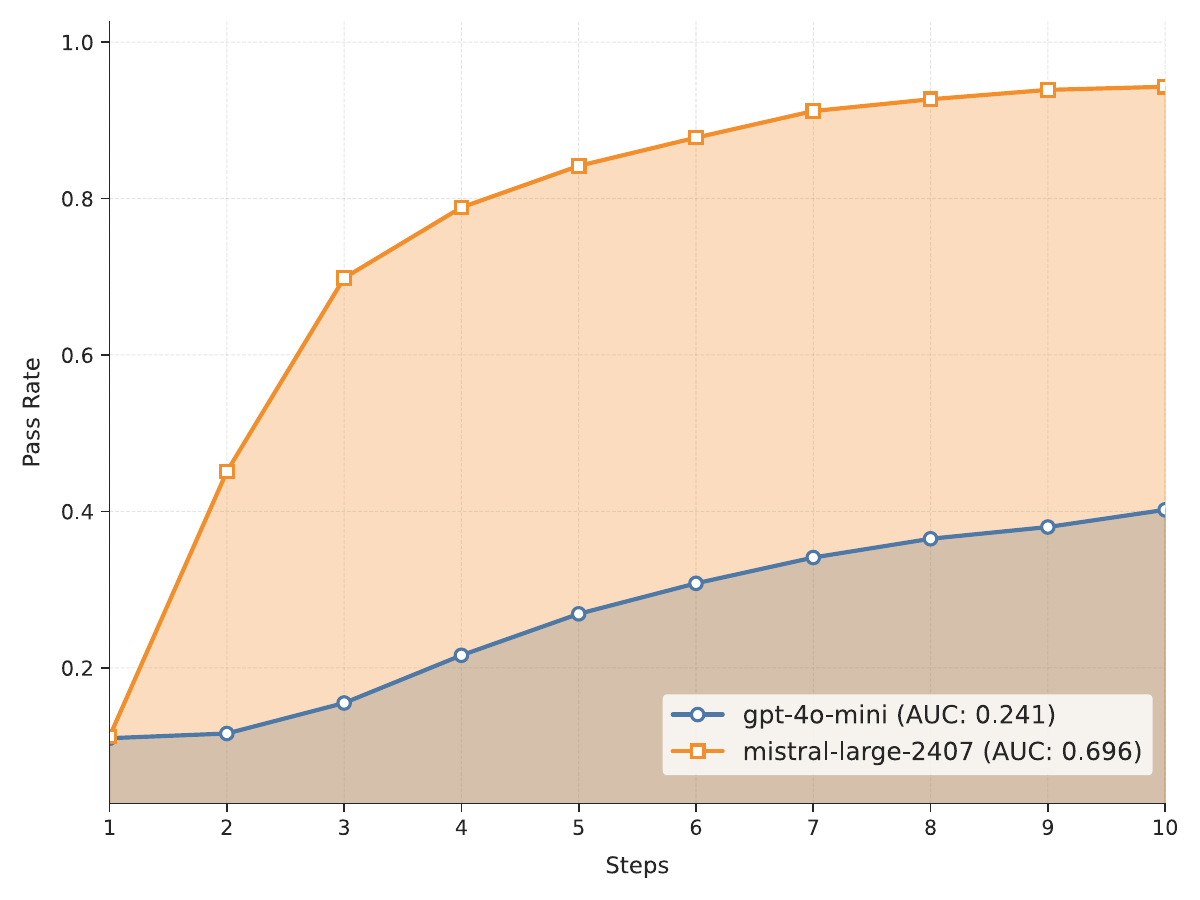}}
    \subfloat[Average Pass Steps over Steps]{
    \includegraphics[width=0.5\columnwidth]{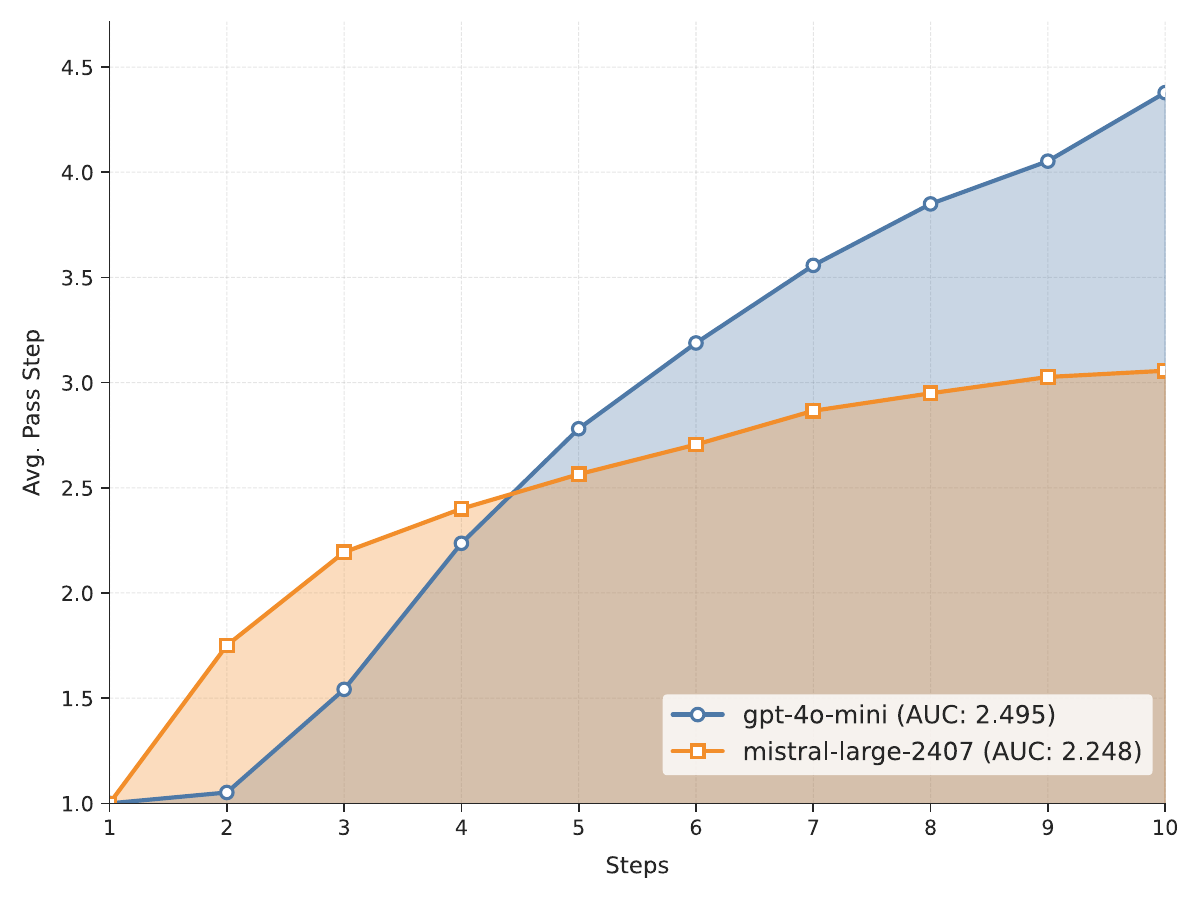}}
    \caption{Step-wise performance metrics comparing GPT-4o-mini and Mistral Large 2 models. (a) Pass Rate over Steps illustrates the cumulative success rate in fixing misconfigurations across repair attempts. (b) Average Pass Steps over Steps indicates the efficiency of repair processes, where lower values represent more efficient repairs.}

    \label{fig:prs_apss}
\end{figure}

\textbf{Results:} The results, presented in Table \ref{table:result} and visualised in Fig. \ref{fig:prs_apss}, demonstrate both the potential and current limitations of LLM-based configuration repair.

\begin{table*}[!htb]
\centering
\begin{tabular}{|l|c|c|c|c|c|c|c|}
\hline
LLM                 & PR(\%) & PSR(\%) & APS  & AUC$_{\text{PRS}}$ & AUC$_{\text{APSS}}$ & Sec Imp & Avg. Introduced Err \\
\hline
GPT-4o-mini        & 40.2  & 99.8   & 4.38 & 0.241              & 2.495               & 0.906   & 0.029              \\
Mistral Large 2    & 94.3  & 100    & 3.06 & 0.696              & 2.249               & 0.986   & 0.024              \\
\hline
\end{tabular}
\caption{Comprehensive performance comparison of different Large Language Models (LLMs) in security configuration repair tasks. Metrics include Pass Rate (PR), Parse Success Rate (PSR), Average Pass Steps (APS), Area Under the Curve metrics (AUC$_{\text{PRS}}$ and AUC$_{\text{APSS}}$), Security Improvement (Sec Imp), and Average Introduced Errors (Avg. Introduced Err).}
\label{table:result}
\end{table*}

GPT-4o-mini exhibited considerable limitations in its ability to correct misconfigurations effectively, achieving only a 40.2\% PR and 99.8\% PSR. This relatively low success rate was accompanied by decreasing efficiency over time, as indicated by an increasing APS that reached 4.38 steps. The model showed a concerning tendency to introduce new errors, with an error introduction rate of 0.029. In contrast, Mistral Large 2 demonstrated remarkably strong capabilities, achieving a 94.3\% PR and 100\% PSR. This high success rate was complemented by consistent efficiency, maintaining 3.06 APS, whilst demonstrating strong security improvement (0.986) with a low rate of introduced errors (0.024), suggesting its potential suitability for production environments.

The step-wise analysis can be better understood through the curves in Fig. \ref{fig:prs_apss}. The Pass Rate over Steps curve in Fig. \ref{fig:prs_apss}(a) shows the cumulative success rate as repair attempts progress, with the Area Under Curve ($\text{AUC}_{\text{PRS}}$) quantifying the overall repair effectiveness, a higher value indicates better overall success (Mistral: 0.696 vs GPT-4o-mini: 0.241). Conversely, Fig. \ref{fig:prs_apss}(b) displays the Average Pass Steps required at each step limit, where the Area Under Curve ($\text{AUC}_{\text{APSS}}$) measures repair efficiency, lower values indicate more efficient repairs (Mistral: 2.249 vs GPT-4o-mini: 2.495), demonstrating Mistral Large 2's superior performance in both dimensions.

\begin{tcolorbox}[
    colback=white,
    colframe=black,
    boxrule=0.5pt,
    arc=2pt,
    left=10pt,
    right=10pt,
    top=10pt,
    bottom=10pt,
    width=\columnwidth
]
\textbf{RQ1 Summary.} LLMs are promising for automated configuration repair. Mistral Large 2 achieved superior performance (94.3\% PR, 100\% PSR, 3.06 APS) compared to GPT-4o-mini (40.2\% PR, 99.8\% PSR, 4.38 APS). Both models maintained low error rates ($<$0.03), demonstrating viability for production use.
\end{tcolorbox}

\subsection{\textbf{RQ2: What misconfiguration types are hard to correct?}}
\textbf{Motivation:} Understanding the limitations of LLM-based correction is essential for both practical implementation and future research directions. We analyse patterns in failed correction attempts and identify categories of misconfigurations that consistently challenge LLM-based correction. By examining the relationships between configuration complexity and correction success, we can identify edge cases that require additional mechanisms or human intervention. This analysis of challenging cases can provide valuable insights for future improvements in automated configuration correction systems and help establish realistic expectations for their capabilities.

\textbf{Method:} According to the intermediate and final results we logged in RQ1, after counting all errors' categories of input files and output files, we obtained and analysed each category's change (fixed or not for each category).

\begin{figure*}[!htb]
    \centering
    \subfloat[GPT-4o-mini\label{fig:compare_4o}]{
        \includegraphics[width=0.99\textwidth]{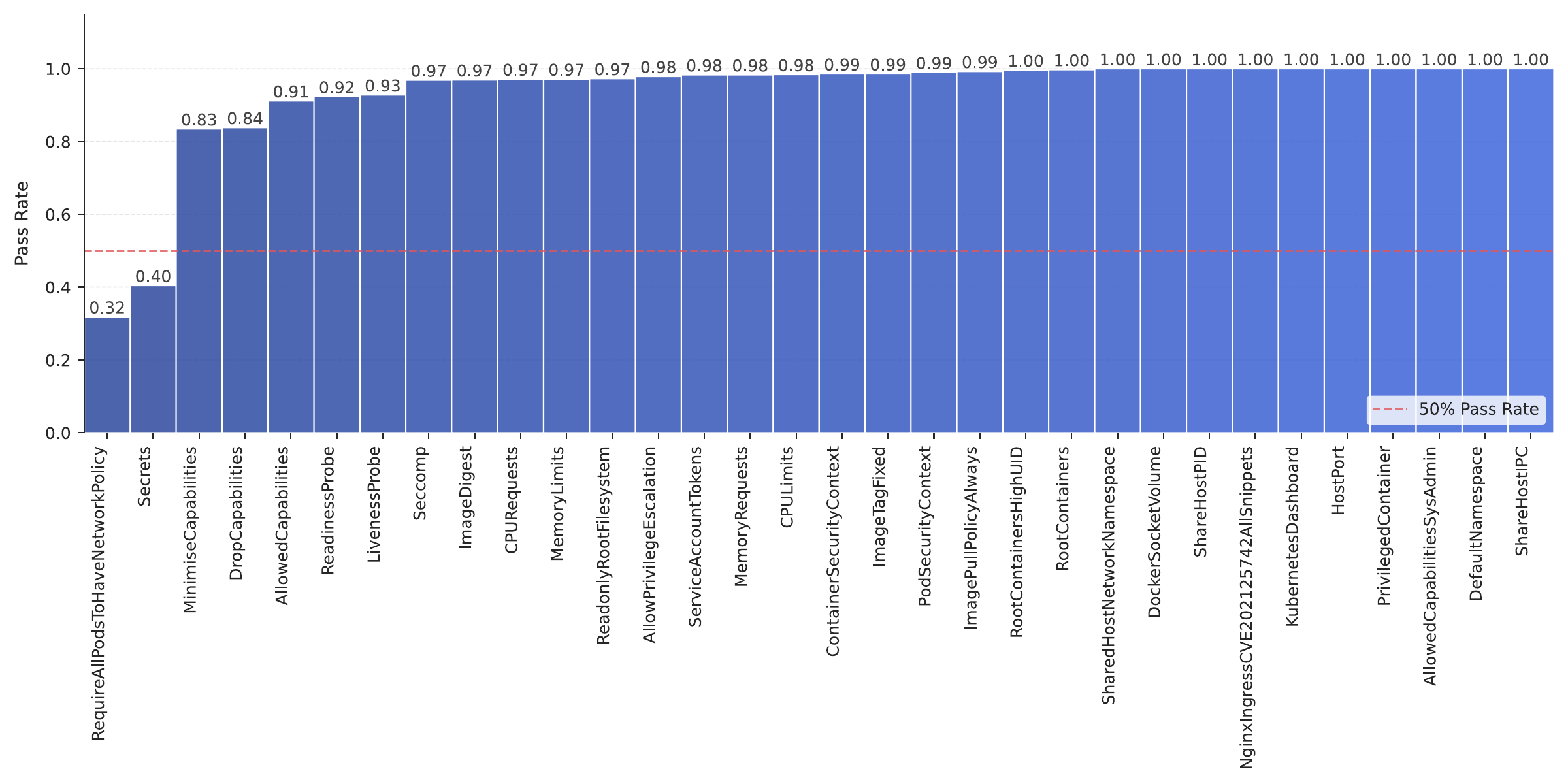}}
    \hfill
    \subfloat[Mistral Large 2\label{fig:compare_mis}]{
        \includegraphics[width=0.99\textwidth]{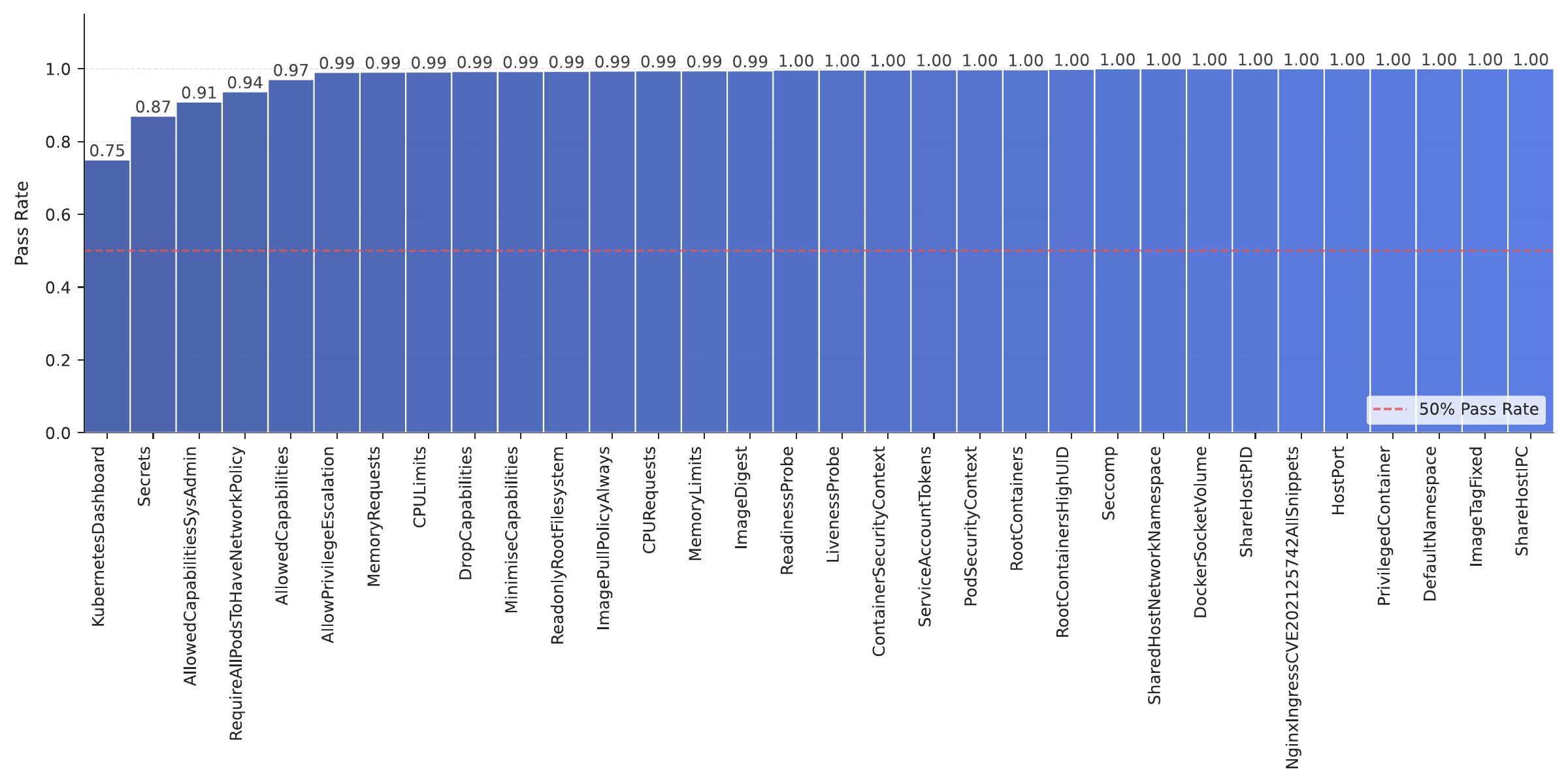}}
    \caption{Performance comparison of different Large Language Models (LLMs) (GPT-4o-mini and Mistral Large 2) in fixing Kubernetes security misconfigurations, categorised by policy type. The results demonstrate varying success rates across different security policy categories, with Mistral Large 2 consistently showing higher performance across most categories.}
    \label{fig:compare}
\end{figure*}

\textbf{Results:} Our analysis across different security policy types revealed distinct patterns in the difficulty of correcting various misconfigurations. As shown in Fig. \ref{fig:compare}, the success rates varied significantly depending on both the type of security policy and the model's capabilities.

Basic security configurations, particularly those involving resource limits and simple networking policies, consistently showed high success rates across all models. These configurations typically involve straightforward parameter adjustments and have clear, well-defined correction paths. The high success rates in these cases suggest that LLMs can effectively handle routine security configuration tasks that form the bulk of day-to-day security maintenance.

However, our analysis identified several categories of misconfigurations that proved consistently challenging to correct. Complex security contexts, particularly those involving privilege-related configurations, posed significant challenges for GPT-4o-mini, with Pass Rates falling below 50\%. These cases often involve intricate interdependencies between different security parameters and require a deeper understanding of security implications.

While Mistral Large 2 showed better performance across all categories, it still encountered occasional difficulties with advanced security policies, particularly those involving complex network configurations.
These configuration types usually have limited data in the wild (see Fig.~\ref{fig:data_distro}), and such data imbalance has been shown to impede the performance of vulnerability prediction models~\cite{le2024mitigating}.
We have also shown that even advanced models may require additional support mechanisms when handling such highly specialised security configurations.

\begin{tcolorbox}[
    colback=white,
    colframe=black,
    boxrule=0.5pt,
    arc=2pt,
    left=10pt,
    right=10pt,
    top=10pt,
    bottom=10pt,
    width=\columnwidth
]
\textbf{RQ2 Summary.} Basic security configurations showed consistently high PR across models. Complex security contexts, especially privilege-related configurations, proved challenging for GPT-4o-mini with $<$50\% PR. While Mistral Large 2 demonstrated a better performance across different policy types, network configurations remained challenging for this model.
\end{tcolorbox}

\subsection{\textbf{RQ3: What context types are most useful for correction?}}
\textbf{Motivation:} Given our framework's integration of RAG and prompting techniques, understanding the impact of different contextual information is crucial. We plan to explore the relative importance of various context sources in the correction process, including error messages from SATs, security documentation and best practices, and configuration file structure and metadata. This investigation helps establish optimal approaches for combining domain-specific knowledge with LLMs' capabilities in security-critical applications.

\textbf{Method:} Our ablation study was conducted using Mistral Large 2 due to its strong performance demonstrated in RQ1. While all parameters and settings remained identical with RQ1, we only varied the context retrieval method. There are three types of contexts in Checkov: scanning results, source code and Prisma documents from URL. The raw scanning results are mandatory for each run, so we tested four variants: scanning results only, scanning results with source code, scanning results with Prisma document content, and full context combining all three sources as used in RQ1.

\begin{table*}[!htb]
\centering
\begin{tabular}{|l|c|c|c|c|c|c|c|}
\hline
Context Type          & PR(\%) & PSR(\%) & APS & AUC$_{\text{PRS}}$ & AUC$_{\text{APSS}}$ & Sec Imp  & Avg. Introduced Err \\
\hline
ckv\_out              & 88.0  & 95.9   & 3.64 & 0.598              & 2.531               & 0.9926   & 0                  \\
ckv\_out + code       & 90.3  & 92.2   & 2.68 & 0.701              & 2.105               & 0.9946   & 0.0134             \\
ckv\_out + prisma     & 65.2  & 98.4   & 4.29 & 0.399              & 2.582               & 0.9675   & 0.0031             \\
\hline
ckv\_out + code + prisma (RQ1)   & 94.3  & 100    & 3.06 & 0.696              & 2.249               & 0.986   & 0.024              \\
\hline
\end{tabular}
\caption{Comprehensive performance comparison of different context types in security configuration repair tasks.
Metrics include Pass Rate (PR), Parse Success Rate (PSR), Average Pass Steps (APS), Area Under the Curve metrics (AUC$_{\text{PRS}}$ and AUC$_{\text{APSS}}$), Security Improvement (Sec Imp), and Average Introduced Errors (Avg. Introduced Err).}
\label{table:ablation}
\end{table*}

\begin{figure}[!htb]
    \centering
    \subfloat[Pass Rate over Steps]{
    \includegraphics[width=0.5\columnwidth]{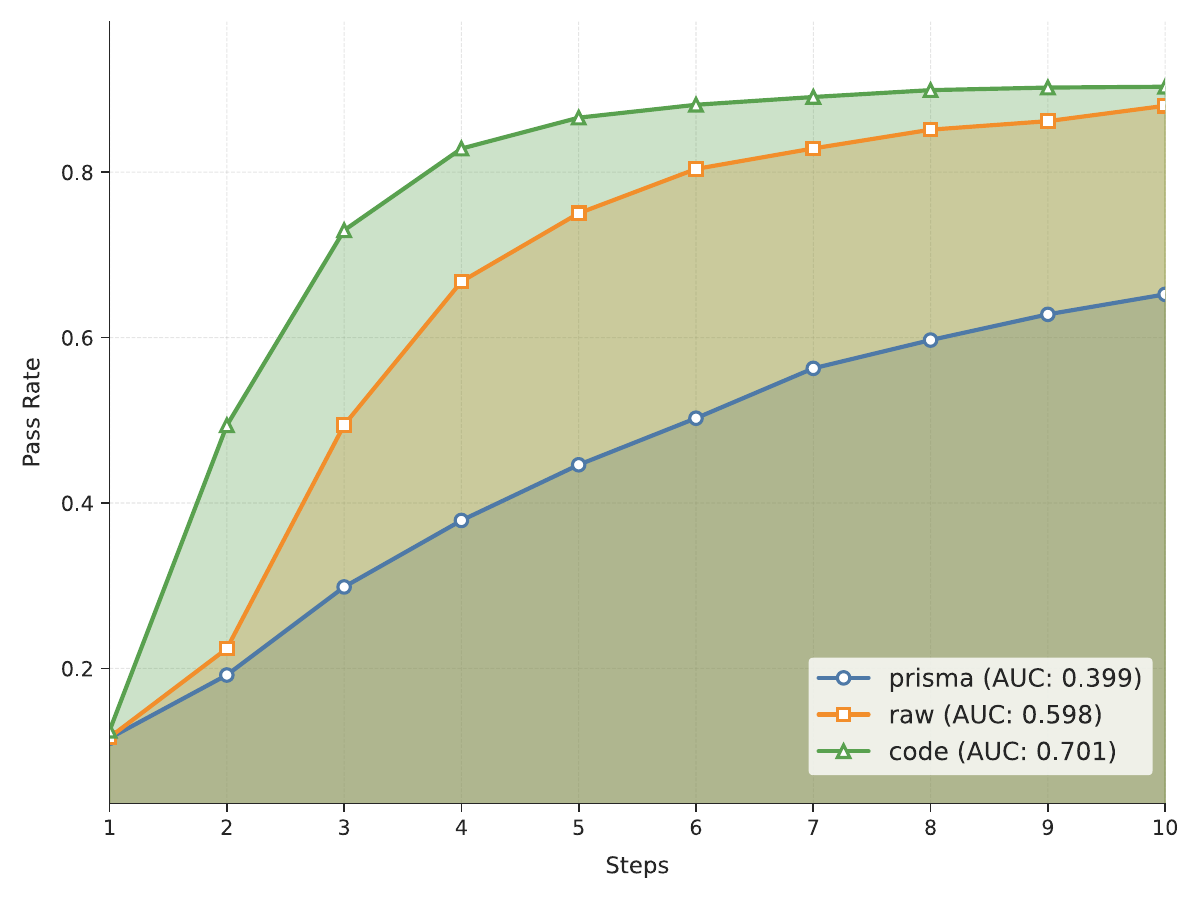}}
    \subfloat[Average Pass Steps over Steps]{
    \includegraphics[width=0.5\columnwidth]{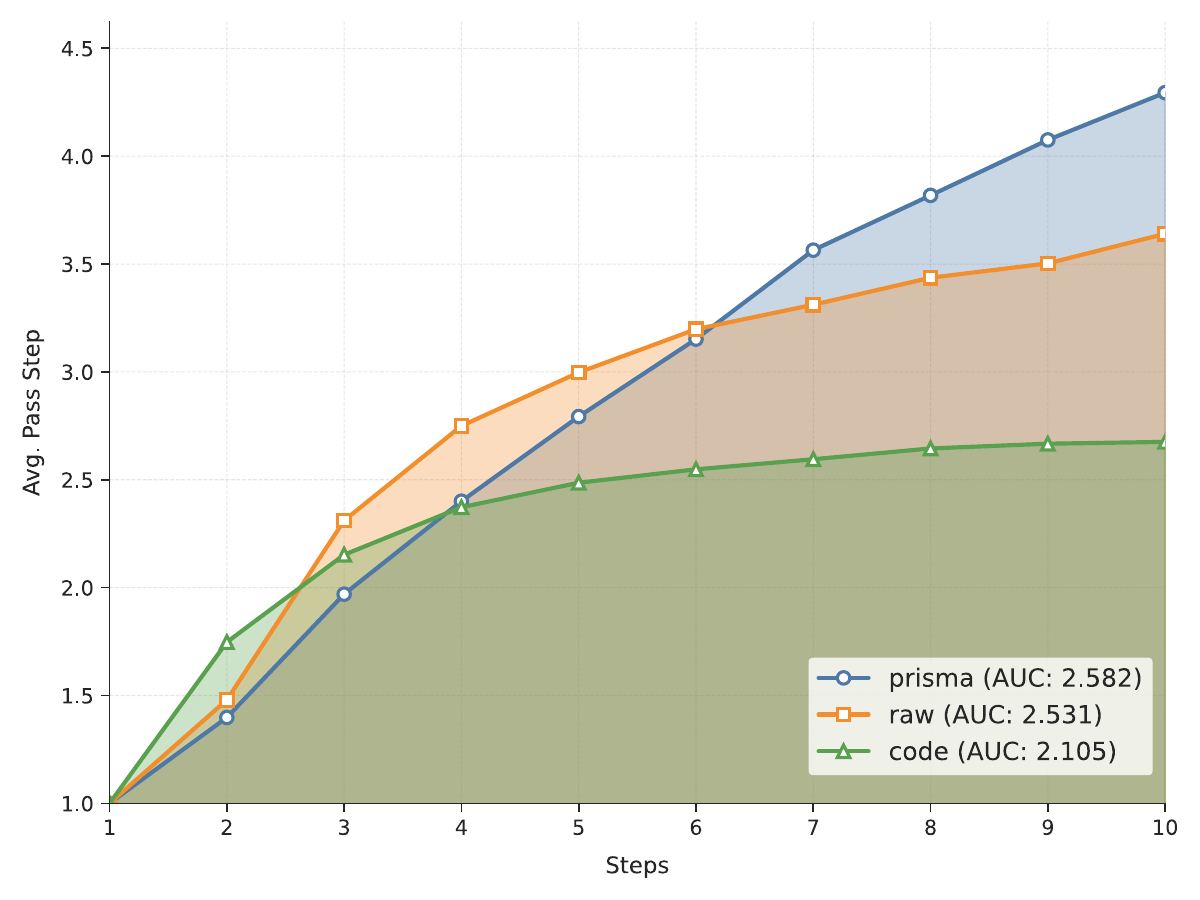}}
    \caption{Ablation study results comparing different context types in the repair process. The graphs display (a) Pass Rate progression and (b) Average Pass Steps across repair iterations for three context configurations: base configuration, source code-enhanced, and Prisma documentation-enhanced. Results demonstrate the superior performance of the source code-enhanced context in both repair success rate and efficiency.}
    \label{fig:abl_prs_apss}
\end{figure}

\textbf{Results:} The results, detailed in Table \ref{table:ablation} and visualised in Fig. \ref{fig:abl_prs_apss}, provide important insights into the relative value of different context types in the repair process.

The base configuration, utilising only Checkov output ($\text{ckv}_{\text{out}}$), demonstrated balanced performance with a PSR of 95.9\% and a PR of 88.0\%. This relatively strong performance suggests that basic error messages provide essential information for many repairs without introducing excessive complexity. The $\text{AUC}_{\text{PRS}}$ value of 0.598 indicates consistent performance across multiple repair attempts.

When enhanced with source code context ($\text{ckv}_{\text{out}}+\text{code}$), the system showed improvement in several metrics. The PR increased to 90.3\%, while the $\text{AUC}_{\text{PRS}}$ improved to 0.701, indicating both higher success rates and more consistent performance. Most notably, the APS decreased to 2.68, suggesting that source code context enables more efficient repairs. This improvement likely stems from the precise validation logic and expected values provided by the source code.

Interestingly, the addition of Prisma documentation alone ($\text{ckv}_{\text{out}}+\text{prisma}$) produced lower performance compared to other configurations. While achieving a high PSR at 98.4\%, the overall PR dropped significantly to 65.2\%, with reduced efficiency (APS: 4.29). However, when combined with both Checkov output, source code and Prisma documentation (full context as used in RQ1), the framework achieved its best overall performance refer to RQ1. This suggests that while documentation alone may introduce noise, its combination with precise technical context creates a complementary effect, providing both the detailed technical requirements from source code and the broader security context from the documentation. The full context configuration achieved superior performance across all metrics, demonstrating that the combination of all three context types provides the most robust foundation for repair generation when properly integrated.

\begin{tcolorbox}[
    colback=white,
    colframe=black,
    boxrule=0.5pt,
    arc=2pt,
    left=10pt,
    right=10pt,
    top=10pt,
    bottom=10pt,
    width=\columnwidth
]
\textbf{RQ3 Summary.} Source code context provided the most effective guidance (90.3\% PR), while basic Checkov output alone achieved 88.0\% PR. Adding Prisma documentation decreased performance (65.2\% PR), indicating that precise, technical context outperforms comprehensive documentation for repairs.
\end{tcolorbox}

\section{Discussion}

\subsection{LLMSecConfig and Beyond}
Our experimental results reveal important insights about automated security configuration repair in container environments, while also highlighting areas for further development. 

The significant performance gap between Mistral Large 2 and GPT-4o-mini demonstrates that advanced model architectures are crucial for processing complex security contexts with interdependent settings. This difference is particularly evident in the models' ability to maintain consistency across diverse policy types while preserving existing configuration intent.

Our ablation study reveals insights about context utilisation in repair generation for container misconfigurations. Source code context consistently provides the most effective guidance, achieving a 90.3\% PR. While comprehensive, documentation can add noise that reduces repair efficiency. This suggests that future implementations should prioritise precise, structured context over verbose documentation, particularly for models with limited processing capacity.

Organizations adopting automated configuration repair should carefully consider their model selection strategy based on our experimental results. The demonstrated success of open-source models like Mistral Large 2, achieving a 94.3\% pass rate while maintaining a low error introduction rate of 0.024, suggests their viability for production environments. However, the complexity of certain configuration types, particularly in advanced network policies, indicates that human oversight remains valuable for high-risk scenarios.

The strong performance of source code-based context retrieval, evidenced by its 90.3\% PR compared to documentation-based approaches at 65.2\%, emphasises the importance of maintaining well-structured policy implementations. Organizations should focus on establishing clear mappings between security checks and their underlying validation logic to maximise repair effectiveness. Furthermore, our successful implementation of multi-stage validation, achieving a 100\% PSR while maintaining security compliance, demonstrates the necessity of comprehensive validation pipelines in production deployments. These findings collectively provide a practical foundation for implementing automated security configuration repair while highlighting specific areas that require careful consideration during deployment.

\subsection{Threats to Validity}
A primary internal validity threat lies in the potential suboptimal configuration of our framework and baseline models. While it is impractical to explore all, we mitigated this by evaluating multiple parameter settings and adapting best practices from relevant literature for LLM-based repair tasks. We also conducted extensive sensitivity analysis to understand the impact of key parameters on repair performance. We choose 0.5 as LLMs Temperature after experiments to find a balance between stabilisation and usability of retry mechanism.

For external validity, our findings may not generalise to all CO platforms and security scenarios as it was impractical to identify all security issues in the wild~\cite{le2024latent}. We addressed this threat by evaluating our approach on 1,000 real-world Kubernetes configurations from ArtifactHub, covering diverse project scales and security requirements. While focusing on Kubernetes, our dataset spans various deployment patterns and security policies commonly found in production environments.

\section{Related Work}
\subsection{Configuration Analysis and Management}
Recent research has made significant progress in understanding and addressing container security challenges through various methodological approaches. Static analysis techniques have been a commonly used approach, with studies by Smith et al.\cite{6987569} and Johnson et al.\cite{4534479} developing tools for analysing container configurations without execution. These approaches primarily focus on identifying potential security issues through pattern matching and semantic analysis, providing early detection of vulnerabilities in configuration specifications.

Complementing static analysis, dynamic analysis methods have been developed to provide runtime security monitoring and enforcement. Sultan et al.\cite{sultan2019container} proposed a comprehensive framework for detecting and preventing security violations during container execution, enabling real-time protection against emerging threats. Their work demonstrates the importance of runtime monitoring in maintaining container security, particularly in dynamic deployment environments.

The application of Machine Learning (ML) techniques to configuration security represents a more recent development in the field. Studies by Bandari \cite{bandari2021comprehensive} and Farkouh et al.\cite{farkouh2023intelligent} have demonstrated the potential of ML approaches in identifying patterns within secure and vulnerable configurations. These data-driven approaches offer promising capabilities for automating security analysis, though they often require substantial training data and careful feature engineering.

KGSecConfig\cite{haque2022kgsecconfig} introduced a structured approach to configuration security, which leverages knowledge graphs and ontologies to represent and reason about security configurations. This semantic approach enables a more sophisticated analysis of configuration relationships and dependencies, though it requires careful knowledge of engineering and maintenance.

While these approaches have advanced our understanding of container security analysis, they primarily focus on detection rather than automated repair. Our work extends beyond detection by introducing an end-to-end framework that not only identifies misconfigurations but also automatically generates and validates fixes, addressing a critical gap in existing container security management solutions.

\subsection{Source Code Vulnerability Detection and Fixing}
The domain of automated vulnerability detection and repair in source code provides valuable insights for configuration security management. Traditional approaches have relied heavily on SATs and pattern matching. For example, Fabian et al.\cite{yamaguchi2014modeling} introduced a code property graph for vulnerability detection, combining abstract syntax trees, control flow graphs, and program dependence graphs to identify complex vulnerability patterns.
Building on this foundation, there have been major advances in AI (Machine Learning, Deep Learning, and recently LLMs) for detecting and addressing vulnerabilities in source code~\cite{Harer2018AutomatedSV,le2021deepcva,chakraborty2021deep,le2022survey,le2022towards,fu2022linevul,le2022use,fu2022vulrepair,nguyen2024automated,le2024software}. These prior studies have established key principles about context utilisation and fix validation that parallel our approach to configuration repair.

While code and configuration vulnerabilities present distinct challenges, the underlying principles of combining traditional analysis tools with AI capabilities remain valuable. Our work adapts these lessons to the specific requirements of container configurations, addressing unique challenges such as security parameter interdependencies and the need to maintain operational stability while enhancing security posture.

\section{Conclusion}
This paper presents a novel framework that combines SATs with LLMs for automated security configuration repair in COs. The framework's success ($\approx$94\% repair accuracy), particularly with context-aware repair generation, shows the potential of integrating traditional static analysis with modern AI techniques. This work represents a significant step toward automated, secure container deployment management.

\section{Data Availability}
\label{sec:data_avail}
The data and code of this study are available at \url{https://figshare.com/s/2a9be8ccfbec9d8ba199}.

\bibliographystyle{IEEEtran}
\bibliography{IEEEabrv,submission}

\end{document}